\documentclass[journal,12pt,onecolumn]{IEEEtran}

\usepackage{mathrsfs}
\usepackage{amsfonts}
\usepackage{tablefootnote}
\usepackage{algpseudocode}
\usepackage[linesnumbered, ruled]{algorithm2e}
\usepackage{amsmath}
\usepackage{url}
\usepackage{makecell}
\usepackage{graphicx}
\usepackage{textcomp}
\usepackage{balance}
\usepackage[table,xcdraw]{xcolor}
\usepackage{booktabs}
\usepackage{multirow}
\usepackage{soul,color}
\usepackage{array}
\newcolumntype{C}[1]{>{\centering\let\newline\\\arraybackslash\hspace{0pt}}m{#1}}
\usepackage{hyperref}

\usepackage{cite}
\usepackage{subcaption} 

\begin{document}

\title{Benchmarking Deep Learning Frameworks for Automated Diagnosis of Ocular Toxoplasmosis: A Comprehensive Approach to Classification and Segmentation}

\author{Syed~Samiul~Alam \emph{Student Member, IEEE} \textsuperscript{1, ‡}, Samiul~Based~Shuvo\textsuperscript{2, ‡}, Shams~Nafisa~Ali\textsuperscript{2}, Fardeen~Ahmed\textsuperscript{3}, Arbil~Chakma \emph{Student Member, IEEE}\textsuperscript{1}, Yeong~Min~Jang \emph{Member, IEEE}\textsuperscript{1, *}
\thanks{\textsuperscript{‡}These authors contributed equally and share first authorship.}
\thanks{\textsuperscript{*}Corresponding author: Yeong~Min~Jang (e-mail: yjang@kookmin.ac.kr).}
\thanks{\textsuperscript{1}Syed~Samiul~Alam, Arbil~Chakma, and Yeong~Min~Jang are with Department of Electronics Engineering, Kookmin University, Seoul 02707, South Korea.}
\thanks{\textsuperscript{2}Samiul~Based~Shuvo, Shams~Nafisa~Ali are with Department of Biomedical Engineering, Bangladesh University of Engineering and Technology (BUET), Dhaka-1205, Bangladesh.}
\thanks{\textsuperscript{3}Fardeen~Ahmed is with AIMS Lab, Institute of Research, Innovation, Incubation and Commercialization (IRIIC), United International University (UIU), Dhaka 1212, Bangladesh and Department of Biomedical Engineering, McKelvey School of Engineering, Washington University in St. Louis, Brookings Dr, St. Louis, USA.}

}


\maketitle
\begin{abstract}
Ocular Toxoplasmosis (OT), is a common eye infection caused by T. gondii that can cause vision problems. Diagnosis is typically done through a clinical examination and imaging, but these methods can be complicated and costly, requiring trained personnel. Lately, the use of artificial intelligence to diagnose different ocular diseases by analysing fundus images has been gaining traction. Despite that, there has not been much work done focusing on the detection of OT. To address this issue, we have created a benchmark study that evaluates the effectiveness of existing pre-trained  networks using transfer learning techniques to detect OT from fundus images. Furthermore, we have also analysed the performance of transfer-learning based segmentation networks to segment lesions in the images. This research seeks to provide a guide for future researchers looking to utilise DL techniques and develop a cheap, automated, easy-to-use, and accurate diagnostic method. We have performed in-depth analysis of different feature extraction techniques in order to find the most optimal one for OT classification and segmentation of lesions. For classification tasks, we have evaluated pre-trained models such as VGG16, MobileNetV2, InceptionV3, ResNet50, and DenseNet121 models. Among them, MobileNetV2 outperformed all other models in terms of Accuracy (Acc), Recall, and F1 Score outperforming the second-best model, InceptionV3 by 0.7\% higher Acc. However, DenseNet121 achieved the best result in terms of Precision, which was 0.1\% higher than MobileNetV2. For the segmentation task, this work has exploited U-Net architecture. In order to utilize transfer learning the encoder block of the traditional U-Net was replaced by MobileNetV2, InceptionV3, ResNet34, and VGG16 to evaluate  different architectures moreover two different two different loss functions (Dice loss and Jaccard loss) were exploited in order to find the most optimal one. The MobileNetV2/U-Net outperformed ResNet34 by 0.5\% and 2.1\% in terms of Acc and Dice Score, respectively when Jaccard loss function is employed during the training.
\end{abstract}

        


%
\IEEEpeerreviewmaketitle

\section{Introduction}
Ocular toxoplasmosis (OT) is an eye infection caused by the  \textit{Toxoplasma gondii (T. gondii)}, an obligate intracellular protozoan parasite. Globally, this disease is the most common cause of infection in the posterior segment of the eye \cite{delair2011clinical,bowling2016kanski}, with approximately one-third of humans estimated to be chronically infected with \textit{T. gondii} \cite{subauste2011review,tenter2000toxoplasma}. However, this prevalence is not constant across all geographic regions and may vary depending on the regions' weather, population's diet, and hygiene maintenance \cite{bonfiolitoxoplasmosis,garweg2005determinants,tugal2005active,remington2004recent,woods1954study}. 

The most common presenting symptoms are a unilateral decrease in visual acuity, sudden onset of floaters, hazy vision, photophobia and a pigmented chorioretinal (affecting the choroid or retina of the eye) scar\cite{small2019congenital,melamed2010ocular}. The disease can also present as a white inflammation of the retina, and of the vitreous humour. In most cases, OT is diagnosed clinically by an opthalmologist based on the appearance of the characteristic lesion during an eye examination. The eye examination consists of a series of tests including those for visual acuity, pupil function, extraocular muscle motility; in case of OT, the examination also includes opthalmoscopy (also known as fundoscopy), where an instrument called the ophthalmoscope is used to project light rays through the pupil to examine the fundus of the eye (fundus), consisting of the retina, optic disc, choroid, and blood vessels. In atypical cases, ocular fluid is extracted and processed using polymerase chain reaction (PCR) to detect parasite DNA. However, the PCR possesses the potential for false-positive results due to contaminating DNA in reagents, equipment or non-viable organisms, as well as false-negative results due to defective primer design, sample or reagents \cite{van2003cme}. 

\begin{figure*}[t]
    \centering
    \includegraphics[width=0.85\linewidth]{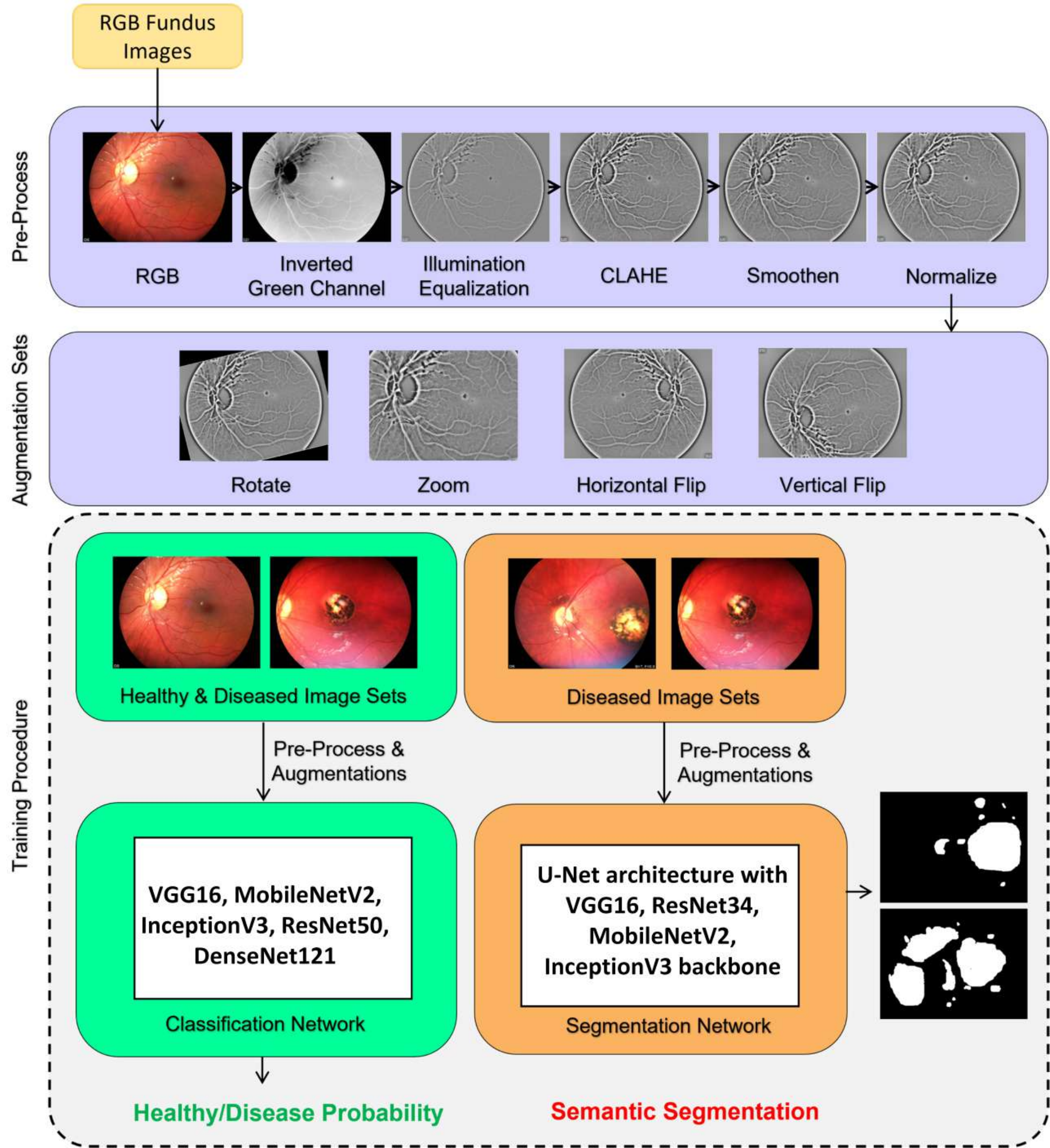}
    \caption{\textbf {RGB fundus images went through several pre-processing steps such as: Inverted green channel, Illumination equalization, CLAHE, Smoothening, and Normalization and augmentation steps (Rotate, zoom, horizontal flip, and vertical flip) before training the binary classification (Healthy and Disease) network and the segmentation network.}}
    \label{f1}
\end{figure*}

The complexities of the current diagnostic procedures for the disease, combined with the costs associated with taking them, the requirement of trained personnel make the disease difficult to detect. Furthermore, the possiblities of a wrong diagnosis adds to the difficulties in testing for OT. Research has been conducted to develop newer, more effective methods for diagnosis \cite{ozgonul2017recent}. Most of these methods attempt to improve existing PCR methods by analyzing different DNA targets in patients
with OT. These studies are focused on Toxoplasma B1 gene, a genomic fragment found in the DNA \cite{steeples2016real, sugita2011diagnosis}. \cite{fekkar2008comparison} reported that combining the PCR test with a calculation of the Goldmann-Witmer coefficient (GWC) \cite{goldmann1954antikorper}, which compares the Toxoplasma - specific antibodies in ocular fluids and in serum, significantly improves diagnostic sensitivity of the disease. While these advancements help in increasing the diagnostic Accuracy (Acc) of the disease, the issue of overcomplexity, high costs and requirement of advanced resources and personnel still remain. In the age of AI, tremendous progress has been achieved in leveraging AI across a range of industries, including disease diagnosis, noise reduction, and segmentation of diverse medical data in both 1D~\cite{alam2023rf,shuvo2023nrc,shuvo2021cardioxnet,ali2023end,huda2020low} and 2D domains~\cite{gaona2022deep,pranata2019deep,souza2019automatic}. In the same vein, deep learning (DL) techniques have been employed to analyze fundus images for diagnosis of various ocular diseases such as glaucoma \cite{ahn2018deep, abbas2017glaucoma, gegundez2021new}, cataracts \cite{dong2017classification, junayed2021cataractnet, raza2021classification}, Diabetic Retinopathy (DR) \cite{gulshan2016development, raman2019fundus}, and age-related macular degeneration (AMD) \cite{peng2019deepseenet}, showing promising results. There has also been several literature studying the use of DL based segmentation networks that focus on lesions found in fundus images \cite{guo2020bin, guo2020lesion, playout2019novel, xue2019deep}. Therefore, it can be established that using DL to detect the presence of diseases in fundus images is a reasonable solution.~\cite{abeyrathna2020directed} proposed clustering-based segmentation for Toxoplasmosis Fundus Images, while the article~\cite{chakravarthy2019approach} explored automatic detection of Toxoplasmosis using a private dataset. A compresensive overview of previous work has been presented in Table~\ref{related_work}. However, there has not been much work done in the classification and segmentation of OT. So, it is necessary to develop a benchmarking study that can be used by future researchers focusing on the use of DL to analyze for the presence of OT. 

This research aims to provide a benchmark for studying OT and for exploring automated, easy-to-access and reliable diagnostic procedures that can detect the presence of OT in an individual. This research aims to address the clinical need by exploring the use of DL techniques to detect and diagnose lesions found in fundus images. 

The primary contributions of this research are to:
\begin{itemize}

\item \textbf{Feasibility of DL Techniques for Ocular Toxoplasmosis Diagnosis:} This research investigates the feasibility of utilizing DL techniques for the diagnosis of Ocular Toxoplasmosis (OT). It explores the application of DL models in accurately detecting and classifying OT-related abnormalities in retinal fundus images. This analysis aims to provide valuable insights into the effectiveness of DL for OT diagnosis, which has not been extensively explored before.

\item 
\textbf{Evaluation of Preprocessing Techniques:} The study evaluates different preprocessing techniques to optimize feature extraction from retinal fundus images specific to OT. These techniques aim to enhance the quality of the images, improve lesion visibility, and enable accurate analysis by DL models. By assessing the impact of various preprocessing methods on the performance of DL models in OT diagnosis, the research contributes to identifying effective strategies for image enhancement and feature extraction.

\item 
\textbf{Benchmarking of DL Networks:} The research benchmarks the effectiveness of DL networks, such as VGG16, DenseNet121, InceptionV3, MobileNetV2, and ResNet50, for diagnosing OT by leveraging transfer learning and pre-trained models, the study compares the performance of these DL networks in accurately identifying and classifying OT-related abnormalities in retinal fundus images. This benchmarking provides valuable insights into the suitability and performance of different DL models for OT diagnosis.

\item 
\textbf{Evaluation of DL Models for Lesion Segmentation:} The research evaluates the performance of DL models in segmenting OT lesions in retinal fundus images. By training and evaluating segmentation models, the study quantifies the accuracy and robustness of DL-based OT lesion segmentation. This contributes to the development of automated methods for precise lesion localization and analysis, aiding in the diagnosis and monitoring of OT.

\item 
\textbf{Consideration of Practical Implications:} The research considers practical implications such as inference time and model complexity in the context of DL-based OT diagnosis. It investigates the computational efficiency of DL models and their suitability for deployment on resource-constrained devices or in real-time scenarios. These considerations provide insights into the practical applicability of DL techniques for OT diagnosis in clinical settings.

\end{itemize}

\begin{table}[!ht]
    
    \caption{Summary of the recent DL based works on analysing fundus images for OT}
    \label{related_work}
    \centering
    \begin{tabular}{p{0.12\linewidth}|p{0.12\linewidth}|p{0.12\linewidth}|p{0.12\linewidth}|p{0.12\linewidth}|p{0.12\linewidth}|p{0.12\linewidth}}
    \hline
        \textbf{Authors}  & \multicolumn{4}{c|}{\textbf{Dataset Details}} & \textbf{Tasks Performed} & \textbf{DL networks used}  \\
        \cline{2-5} & \textbf{Collected from} & \textbf{No. of Images} & \textbf{Publicly available} & \textbf{Ground Truth Segmentation Masks Available} & \\ \hline
        Parra et. al \cite{parra2021automatic} & Hospital de Clínicas and Hospital General Pedriático Acosta Ñu medical centers from Asunción, Paraguay \cite{otfund} & $160$ & \checkmark & \checkmark & Diagnosis & Recurrent Neural Network (RNN) \\ \hline
        Parra et. al \cite{parra2021trust} & Hospital de Clínicas and Hospital General Pedriático Acosta Ñu medical centers from Asunción, Paraguay \cite{otfund} & $160$ & \checkmark & \checkmark & Diagnosis & Convolutional Neural Network (CNN), Transfer Learning using VGG16 and ResNet18\\ \hline
        Hasanreisoglu et. al \cite{hasanreisoglu2020ocular} & $6$ uveitis clinics in Argentina, Turkey, and United States + Control dataset from STARE \cite{stare} & $246+66$ & $\times$ & $\checkmark$ & Diagnosis & Dual Input Hybrid Transfer Learning Model with VGG16\\ \hline
        Chakravarthy et. al \cite{chakravarthy2019approach} & $6$ uveitis clinics in Argentina, Turkey, and United States $+$ Control dataset from STARE \cite{stare} & $246 + 66$ & $\times$ & $\checkmark$ & Diagnosis & CNN\\ \hline
        Abeyrathna et. al \cite{abeyrathna2020directed} & $6$ uveitis clinics in Argentina, Turkey, and United States $+$ Control dataset from STARE \cite{stare} & $246 + 66$ & $\times$ & $\checkmark$ & Segmentation & Mask R-CNN\\ \hline
    \end{tabular}
\end{table}

The paper is structured as follows: Section \ref{mt2} provides an overview of OT disease to lay the foundation for the reader. Then, in Section \ref{mt3}, the datasets used in the study are presented along with their preprocessing approaches. Section \ref{mt4} outlines the various Transfer learning approaches utilized in the study. In Section \ref{mt5}, the benchmark settings used in the experiments are demonstrated, followed by the presentation of the experimental results in Section \ref{mt6}. A comparative discussion of the results is provided in Section \ref{mt7}. In Section \ref{mt9}, the authors outline future research directions and discuss the limitations of their proposed strategy. Finally, the contributions of this work are summarized in Section \ref{mt10}. 

\section{Background}\label{mt2}
An eye infection caused by OT typically affects the retina, causing an individual to suffer various forms vision impairments. The parasite is most commonly transmitted to humans through the fecal-oral route (example: an infected cat litterbox or sandbox), consumption of contaminated food, bodily fluids including blood, and transplant procedures. The disease may also be acquired by a fetus at birth (congenital), from an infected mother through the placenta. Toxoplasm a gondii contracted at birth usually stays dormant, except in situations where the individual's immune system has been compromised, causing the parasite to reactivate, causing ocular pain, blurred vision, and possibly permanent damage, including blindness in infected individuals \cite{glasner1992unusually,holland2000ocular,atmaca2004clinical}. 
After the infection, the majority of individuals present no symptoms at all, but in severe cases, the disease often presents as posterior uveitis (inflammation of the uvea, the middle layer of the eye) with a unilateral chorioretinal lesion and vitritis (cellular infiltration of the vitreous body) \cite{balasundaram2010outbreak}. Upon opthalmoscopic examination, the inflamed vitreous is shown to have a "headlight-in-fog" appearance\cite{dutta2021driving,park2013clinical}. A unifocal area inflammation adjacent to the chorioretinal lesion is also characteristic of the disease \cite{park2013clinical}. Analysing the fundus images of the back of the eye, can therefore, provide significant information pertaining to the presence of OT in a patient.

\begin{figure}[t]
    \centering
    \includegraphics[width=0.5\linewidth]{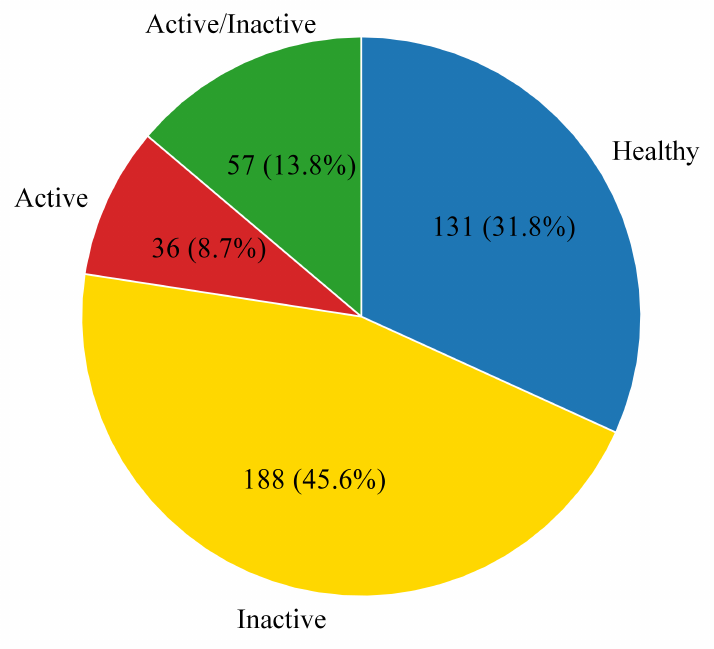}
    \caption{Distribution of data corresponding to the disease classes.}
    \label{pie_dataset}
\end{figure}

\begin{figure*}[h]
  \centering
  \includegraphics[trim=0 0 0 0,clip,width=1\linewidth]{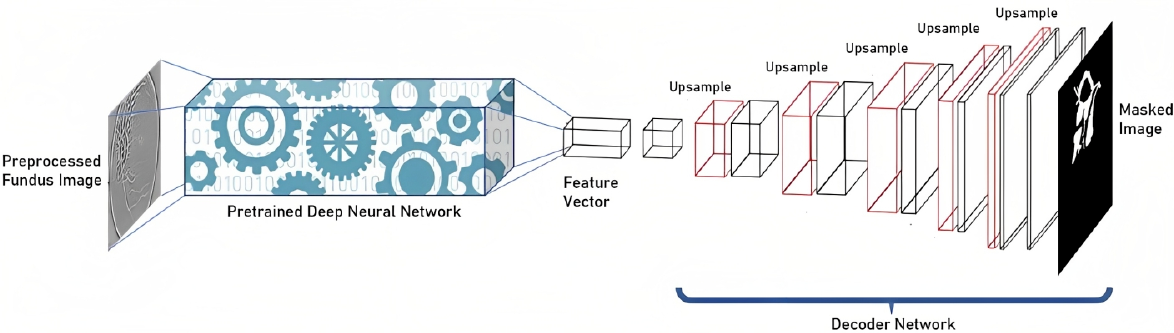}
  \vspace{-5mm}
  \caption{Schematic Layout of the use of transfer learning for Image Segmentation for this research. 
  } 
  \label{fig:10}
\end{figure*}

\section {Methods} \label{mt3}
\subsection{Dataset}
In this work, $412$ fundus images from the publicly available "OT Fundus Images Dataset (OTFID)" have been used~\cite{cardozo2023dataset}. This recently released dataset was collected between $2018$ to $2021$ at two hospital centers: Hospital de Clínicas Medical ($291$ images with a size of $2124 x 2156$ pixels) and Niños de Acosta Ñú General Pediatric Hospital ($121$ images with a size of $1536 x 1152$ pixels) JPG format. These images were of patients with suspected congenital toxoplasmosis infection and were classified into healthy and non-healthy categories. Non-healthy images were further classified as i. inactive only, ii. active only, and iii. active/inactive. The dataset also contains manual segmentation of the images with active and/or inactive lesions. Fig~\ref{pie_dataset} shows the relative distribution of the dataset in each category.


\subsection{Data Preprocessing}
Preprocessing is frequently done on images to reduce the effects like poor illumination, contrast and noise. Across the literature, several types of preprocessing have been mentioned for retinal fundus images~\cite{tsiknakis2021deep, islam2020deep, atwany2022deep}. Some general steps include denoising/smoothening, $\mathscr{N}$, contrast enhancement, illumination equalization, color space transformation etc.~\cite{tsiknakis2021deep}.
\par 
\textbf{Denoising/Smoothing:}
Typically, image data are stored in a compressed form, this makes small diagnostically distinctive structures, infection lesions/scars distorted. Therefore, denoising/smoothing is often considered as a necessary step before passing it to the detection pipeline. A 2D Gaussian filter with standard deviation $\sigma$ and kernel size $k \times k$: 

\begin{equation}
G_{i,j}=\frac{1}{2\pi \sigma^2} \exp\left(-\frac{(i-2k+1)^2+(j-2k+1)^2}{2\sigma^2}\right)
\end{equation}

where $i$ and $j$ are the indices of the kernel element, and $G_{i,j}$ is the corresponding value of the kernel at that location.
\par 
Applying Gaussian filter with a relevant width and variance is one of the most adopted techniques for denoising images because of its ability to  attenuate high-frequency noise while preserving the edges and details of the image~\cite{wu2017automatic, prentavsic2016detection}. Its smoothing effect also acts against random noise by reducing the impact of outliers and small variations in the image~\cite{islam2020deep}. Recently, Non-Local Means Denoising (NLMD) method has been also used for denoising retinal fundus image~\cite{ghosh2017automatic, luo2023deep} as this non-parametric method can not only effectively remove the noise but also preserve the edges and textures of the image~\cite{tsiknakis2021deep}. 

\begin{equation} \label{nlmd}
NLMD(x)(i) = \frac{\sum_{j \in \Omega_i} w(i,j)x(j)}{\sum_{j \in \Omega_i} w(i,j)}
\end{equation}

where $x$ and $i$ denote the input signal and the current pixel location respectively, $\Omega_i$ is the search window around $i$, and $w(i,j)$ is the weight assigned to pixel $j$ based on the similarity between the patches centered at $i$ and $j$. The NLMD method Eqn. (\ref{nlmd}) computes the weighted average of pixels within the search window $\Omega_i$, with the weights given by the similarity between the patches centered at $i$ and $j$. The resulting output signal $NLMD(x)$ represents a denoised version of the input signal $x$.
\par
\begin{algorithm}[t]
\caption{Preprocessing Pipeline}
\label{preprocess_algo}
\textbf{Input:} $X_t = \{x_{t}^{i}\}_{i=1,\dots,n}$ - Fundus images from the dataset \\
\textbf{Output:} $X_{pre} = \{x_{pre}^{i}\}_{i=1,\dots,n}$ - Preprocessed Fundus Images
\begin{enumerate}
    \item Separate the RGB channel intensities from the images, $X_t$ $\rightarrow$ $X_{t}^{r}, X_{t}^{g}, X_{t}^{b}$ 
    \item Invert the Green channel Images, $X_{t}^{g} = 1-X_{t}^{g}$
    \item Apply a mean filter to the green channel, $X_{mf} = X_{t}^{g}*M( x, \hat{x})$
    \item Subtract the background image from the original image, $X_{nb} = X_{t}^{g} - X_mf$
    \item Add the mean intensity of the original image to the corrected image, $X_{corr} = X_{nb} + mean(X_{t}^{g})$
    \item Apply Contrast Limited Adaptive Histogram Equalization (CLAHE), $X_{clahe} = X_{corr} * F_{CLAHE}( y ; \theta)$
    \item Apply the Gaussian filter to the image, $X_{gf} = X_{clahe}*G( x, \hat{x})$
    \item Normalise the resulting Image, $X_{pre} = X_{gf} / max(X_{gf})$
    \item \textbf{return} $X_{pre} = \{x_{pre}^{i}\}_{i=1,\dots,n}$
\end{enumerate}
\end{algorithm}

\textbf{Normalization ($\mathscr{N}$):}
The images standardized to a particular scale so that no bias is present in the data. It is effective in reducing high training times to the network as well. Typically, the normalized image $I_{\text{normalized}}$ at a given pixel location $(x,y)$ is obtained by dividing the original pixel value $I(x,y)$ by the maximum pixel value in the entire image $\max_{x,y} I(x,y)$, as represented in Eqn. (\ref{I_norm}). 

\begin{equation}
\label{I_norm}
I_{\text{$\mathscr{N}$}}(x,y) = \frac{I(x,y)}{\max_{x,y} I(x,y)}
\end{equation}

This $\mathscr{N}$ method scales the pixel values of the image between 0 and 1, making it easier to compare images with different intensity ranges. 

\textbf{Contrast Enhancement:}
To improve the visual quality of images by increasing the contrast between the different regions in an image, contrast enhancement techniques like histogram equalization (HE), contrast limited adaptive histogram equalization (CLAHE), homomorphic filtering, wavelet and retinex methods etc. are employed~\cite{alyoubi2020diabetic, tsiknakis2021deep}. Since CLAHE is a relatively fast algorithm with the feature of local contrast preservation, adjustable contrast, less susceptibility to over-enhancement, it is widely applied in biomedical problems tp increase visibility of the salient parts of an image~\cite{wu2017automatic}.  
\begin{equation}
I_{out}(x,y) =
\begin{cases}
T(x,y) & \text{if } T(x,y) \leq L_{clip} \\
L_{clip} & \text{if } T(x,y) > L_{clip}
\end{cases}
\end{equation}

where $T(x,y)$ is the transformed pixel value obtained by applying the CLAHE algorithm at location $(x,y)$, and $L_{clip}$ is the maximum allowed intensity value after the histogram equalization.

\textbf{Color Space Transformation and Illumination Equalization:}
Since the green channel is rich in information and provides the best background contrast compared to the red (saturated) and blue channels (dark, less valuable information)~\cite{walter2007automatic}, extracting the green channel~\cite{yu2017exudate, tsiknakis2021deep} or using the inverse green channel~\cite{wu2017automatic} for extracting features from medical images have been a long practice. 
\par 
Using the illumination equalization method provided by Hoover et al.~\cite{hoover2003locating} on an inverted green channel image $I_{g}$, the visual appearance of images with non-uniform illumination, shade etc. can be fixed. First, a mean filter of size $k \times k$ is applied to the $I_{g}$ to generate the estimated background image $I_{bg}$ which is then subtracted from
the $I_{g}$ and added to the average intensity $u$ of $I_{g}$ to obtain the illumation equalized image $I_{iq}$.

\begin{equation}
I_{bg} = \text{meanfilter}(I_g, k)
\end{equation}
\begin{equation}
u = \text{mean}(I_g) 
\end{equation}
\begin{equation}
I_{ie} = I_{g} - I_{bg} + u
\end{equation}

In this work, we have experimented with four preprocessing pipelines using the VGG16 architecture for classification on 3-fold data:
\begin{itemize}
    \item CLAHE $\rightarrow$  NLMD $\rightarrow$ $\mathscr{N}$
    \item CLAHE $\rightarrow$ NLMD $\rightarrow$ Gaussian Intensity $\mathscr{N}$ (pixels have a mean of 0 and standard deviation of 1)
    \item Color space transformation (inverted green channel)\textcolor{white}{---}$\rightarrow$ Illumination equalization $\rightarrow$ CLAHE $\rightarrow$ NLMD $\rightarrow$ $\mathscr{N}$
    \item Color space transformation (inverted green channel) 
    \textcolor{white}{---}$\rightarrow$ Illumination equalization $\rightarrow$ CLAHE $\rightarrow$ Gaussian denoising/smoothing ($k = 51$) $\rightarrow$ $\mathscr{N}$
\end{itemize}

From the first two experiments mentioned above, we realized the superiority of the traditional (maximum value based) $\mathscr{N}$ technique over the Gaussian Intensity $\mathscr{N}$ as the letter had lagged by a mean Acc of (4.64$\pm$0.07\%). Then we conducted the last two experiments of the list. Although the existing difference between the methods had negligible difference across all the performance metrics, the processing time for the third method involving NLMD was much longer compared to the last method where Gaussian denoising/smoothing ($k = 51$) has been utilized. Therefore, the last processing pipeline (see Algo.~\ref{preprocess_algo}) has been finalized as the preprocessing pathway for this work to conduct the benchmarking study. 

\begin{algorithm}[h]
    \caption{Augmentation}
    \label{algo_aug}
    \SetKwInOut{Input}{Input}
    \SetKwInOut{Output}{Output}
    \Input{$X_t = \{x_{t}^{i}\}_{i=1,\dots,n}$ and $Y_t = \{y_{t}^{i}\}_{i=1,\dots,n}$ - Fundus image and ground truth mask in the dataset, respectively}
    \Output{$img_{augmented}$ and $mask_{augmented}$}
    Initialize~$img_{augmented}$=$[]$\\
    Initialize~$mask_{augmented}$=$[]$\\
    \For{i $\leftarrow~1~to~\emph{length}(X_t)$}{
        $\theta$ $\gets$ $90$\\
        \For{j $\leftarrow$~$1$~to~\emph{length}($X_t$)}{
            $img_{augmented}$ $\Leftarrow$ $\mathscr{R}(x^{i}, \theta)$\\ 
            $mask_{augmented}$ $\Leftarrow$ $\mathscr{R}(y^{i}, \theta)$\\
            $\theta$ $\leftarrow$ $\theta$+$90$\\
        }
        $img_{augmented}$ $\Leftarrow$ $\mathscr{N}(x^{i})$\\
        $mask_{augmented}$ $\Leftarrow$ $y^{i}$\\  
        $img_{augmented}$ $\Leftarrow$ $\emph{horizontal flip}(x^{i})$\\
        $mask_{augmented}$ $\Leftarrow$ $\emph{horizontal flip}(y^{i})$\\
        $img_{augmented}$ $\Leftarrow$ $\emph{vertical flip}(x^{i})$\\   
        $mask_{augmented}$ $\Leftarrow$ $\emph{vertical flip}(y^{i})$\\  
    }
    \Return ~$img_{augmented}$, $mask_{augmented}$\\
\end{algorithm}

\subsection{Augmentation}
In data augmentation, various transformations such as rotating ($\mathscr{R}$), flipping (horizontal or vertical), shearing, cropping, scaling, jittering of color, hue and brightness, noise addition etc. are applied to the existing data to artificially increase the size and diversity of a dataset. Since the dataset of this work is relatively small and unbalanced, several data augmentation techniques (elaborated at Algo.~\ref{algo_aug}) have been adopted for both classification and segmentation frameworks to enhance the robustness of the model and its ability to generalize new and unseen data. 

\section{Transfer Learning}\label{mt4}
Machine learning and DL applications require vast amounts of data for the training phase of the procedure, which  may not always be available. In such situations, trasnfer learning \cite{tan2018survey} algorithms have proven to be effective solutions for training DL networks. Pre-trained models, models that have already been trained on a sizable dataset like ImageNet \cite{deng2009imagenet}, are utilized in the transfer learning process, where the initial layers of the model are frozen \cite{brock2017freezeout} out, and the final layers of the models are adjusted. Freezing a layer, essentially means that the layer's weight won't and consequently, the layer will not 'learn' anything. Transfer learning is appropriate for this research as the data set is relatively small compared to datasets used to train a full neural network, with a small number of fundus images for all classes. 
\subsubsection{VGG16}
The VGG model, commonly known as VGGNet, is referred to as VGG16. It is a 16-layer convolution neural network (CNN) model. 
In the ImageNet dataset, the VGG16 model can achieve a high test Acc of 92.7\% for 1000 object classes. VGG16 outperforms AlexNet \cite{krizhevsky2017imagenet} (whose first convolutional layer's kernel size is 11x11 and its second layer's kernel size is 5x5) by substituting the large filters with a series of smaller 3x3 filters, which provides the function of a bigger receptive field. Using multiple smaller layers instead of a single large layer means that more non-linear activation layers are present, which enables the network to converge quicker, and therefore, improves the decision functions \cite{simonyan2014very}.

\subsubsection{ResNet}

Residual networks (ResNet) are CNNs designed to address the issue of vanishing gradients in deep neural networks by employing skip connections. ResNet comes in various sizes, such as ResNet34 and ResNet50, which differ in the number of convolutional layers and parameters. ResNet50 has 50 convolutional layers and approximately 11 million parameters. Skip connections are utilized to address the vanishing gradient problem by skipping certain network layers. Each block in the network includes two 3 x 3 convolution layers, batch $\mathscr{N}$, and ReLU activation. ResNet34, on the other hand, has 34 convolutional layers and approximately 21 million parameters. Similar to ResNet50, it uses skip connections, two 3 x 3 convolution layers, batch $\mathscr{N}$, and ReLU activation in each block\cite{he2016deep}. 
\subsubsection{MobileNetV2}
MobileNetV2 is a CNN architecture based on an inverted residual structure, with shortcut connections between narrow bottleneck layers, that is designed to improve mobile and embedded vision systems. A Bottleneck Residual Block is a form of residual block that uses 1x1 convolutions to generate a bottleneck. A bottleneck can be used to reduce the amount of parameters and matrix multiplications. The purpose is to make remaining blocks as tiny as possible so that depth and parameters can be enhanced. The activation function in the model is ReLU. The top layer of the design is a 32-filter convolutional layer, followed by 19 bottleneck levels. \cite{sandler2018mobilenetv2}. 
\subsubsection{InceptionV3}
The InceptionV3 model is an enhanced and optimized version of the InceptionV1 model \cite{szegedy2015going}, which is a deep CNN with increased depth and width that does not consume an excessive computational budget. The InceptionV1 model uses numerous filters of different sizes on the same level to avoid overfitting induced by deep layers of convolutions. Thus, instead of deep layers, parallel layers are used in the inception models, making the model wider rather than deeper.

The Inception model is composed of several Inception modules, with the fundamental module of the InceptionV1 model consisting of four parallel layers: 1x1 convolution, 3x3 convolution, 5x5 convolution, and 3x3 max pooling.

The InceptionV3 model optimized the original network for better model adaption by using numerous strategies such as factorization into smaller convolutions, spatial factorization into asymmetric convolutions, usage of auxiliary classifiers, and efficient grid size reduction. As a result, as compared to the InceptionV1 model, the InceptionV3 model has a deeper network with no loss of speed, improved efficiency, and reduced computational cost \cite{szegedy2016rethinking}.

\subsubsection{DenseNet121}
A DenseNet is a specific kind of CNN that creates dense connections between layers via Dense Blocks. In this kind of network, more specifically in the Dense blocks, all of the layers are connected directly with each other in a feed-forward fashion, and matching the feature-map sizes of each layer. The network has a total of $\frac{L(L+1)}{2}$ direct connections, where L is the number of layers. This is in direct contrast to conventional CNNs, where networks of L layers contains a total of L connections. These L connections in the DenseNet are found between each layer and the layer that comes after it. As inputs for each layer, the feature-maps of all of the preceding layers are used, and the feature-maps of the current layer are used as inputs for all of the subsequent layers. DenseNets have several significant advantages: they solve the problem of vanishing gradients, improve feature propagation, promote feature reuse, and reduce the number of parameters being used\cite{huang2017densely}. DenseNet121 has 120 Convolution layers and 4 Average Pooling layers.
Layers within the same dense block and transition layers, distribute their weights over multiple inputs which allows deeper layers to use previously extracted features.




\begin{algorithm}[t]
\caption{Benchmarking of Pre-trained Classifiers for Diagnosing OT}
\label{Calgo}
\textbf{Input:} $D = \{(x^{(i)}, y^{(i)})\}_{i=1,\dots,N}$ -- dataset of fundus images and corresponding labels, where $x^{(i)}$ is the $i$-th fundus image and $y^{(i)}$ is the corresponding label (0 for healthy and 1 for diseased); $k=5$ -- the number of folds. \\
\textbf{Pre-trained Models:} VGG16, DenseNet121, InceptionV3, MobileNetV2, and ResNet50.\\
\textbf{Loss:} Sparse Categorical Cross-entropy.\\
\textbf{Evaluation Metrics:} Acc, Precision (Pr), Recall (Re), and F1-score (F1).\\
\textbf{Output:} Optimized pre-trained classifier models $M_i$ using 5-fold cross-validation.

\textbf{Steps:}
\begin{enumerate}
  \item Split dataset $D$ into training set $D^{(i)}_{\text{train}}$ and validation set $D^{(i)}_{\text{val}}$ with 80\% and 20\% of data, respectively.
  \item For $i=1$ to $k$, do the following:
    \begin{enumerate}
      \item Define model architecture $M$ with pre-trained weights of VGG16, DenseNet121, InceptionV3, MobileNetV2, or ResNet50.
      \item Compile model $M$ with Sparse Categorical Cross-entropy loss function.
      \item Train $M_i$ on $D^{(i)}_{\text{train}}$ with Sparse Categorical Cross-entropy loss function.
      \item Evaluate model $M_i$ on $D^{(i)}_{\text{val}}$ and record evaluation metrics (Acc, Pr, Re, F1) and save the metrics' values.
    \end{enumerate}
  \item Return $M_{1}, M_{2}, \dots, M_{5}$.
\end{enumerate}
\end{algorithm}

\section{Benchmark settings for diagnosis of OT and OT lesion segmentation}\label{mt5}

The OTFID dataset enables the investigation of diverse issues related to OT, including the segmentation of lesions at the pixel level and classification of the OT condition at the image level. To assess various pre-trained models' performance on this dataset, we designed two tasks. Task 1 delves into the challenging and intriguing research topic of OT classification, while Task 2 applies classic segmentation pre-trained models for medical imaging to numerous OT lesions.

\subsection{Task 1: OT classification}
In Task 1 experiments for this research, all the layers, except the final classification layers of the pre-trained networks being evaluated were frozen, and then connected with 512 dense units of ReLu-activated dense layer and a final classification layer with Softmax activation. The complete OT classification pathway is demonstrated by Algo. ~\ref{Calgo}
\subsection{Task 2: OT lesions segmentation}
For Task 2, segmentation has been used in 'unhealthy' images to segment the lesions as either active, inactive or active/inactive. Fundus images were first input into pre-trained networks for deep-learning feature extraction. The extracted feature vectors were then connected to a decoder \cite{badrinarayanan2017segnet} - a sequence of Up-sampling layers that translates the vectors to data with higher dimensions, to output the segmented image. This framework was based on the U-Net image segmentation model \cite{ronneberger2015u}, the SOTA network for biomedical image segmentation.  The pre-trained networks used for segmentation are VGG16, ResNet34 and MobilenetV2.
The complete OT segmentation pathway is demonstrated by Algo. ~\ref{Salgo}
\subsection{Experimental Setup}

The TensorFlow and Keras frameworks are used to implement the DL architectures. The training and testing of all models are performed on NVidia K80 GPUs and Intel(R) Xeon(R) CPUs. The segmentation network has been trained using either the Dice loss or Jaccard loss function, while the classification networks have been trained using the Sparse categorical cross-entropy function as the loss function.

Both models employ the adaptive learning rate optimizer (Adam) which has a batch size of 128 and the initial learning rate is set at $10^{-4}$. Adam was chosen due to its superior efficacy compared to other optimizers, as demonstrated in previous research~\cite{wilson2017marginal}.

To ensure an equal representation of samples from each class in every batch during the training of the classification model, a mini-batch balancing scheme~\cite{humayun2020towards} is utilized.

A summary of the considered hyperparameters are listed in Table \ref{hyper}.
\begin{table}[!h]
\centering
\caption{Hyper-parameters of the benchmarking framework}
\label{hyper}
\scalebox{1.2}{
\begin{tabular}{@{}C{3.5cm}C{3.2cm}@{}}
\toprule
\hline
Hyperparameters    & Values                 \\ 
\hline
\midrule

Batch size          & $32$                        \\
Learning rate       & $0.0001$                    \\
Epoch               & $200$                        \\
Optimizer           & Adam                      \\
Segmentation network's
Loss function       & Dice loss, Jaccard loss   \\ 
Classification network's
Loss function       & Sparse categorical cross-entropy  \\ 
\hline
\hline
\end{tabular}
}
\end{table}

\begin{algorithm}[h]
\caption{Benchmarking of the pre-trained segmentation network}
\label{Salgo}
\textbf{Input:} $X = \{x^{i}\}_{i=1,\dots,n}$ - fundus images, $Y = \{y^{i}\}_{i=1,\dots,n}$ - binary lesion segmentation masks \\
\textbf{Loss Function:} Dice loss or Jaccard loss\\
\textbf{Evaluation Metrics:} Dice score and IoU\\
\textbf{Output:} $F(x ; \theta)$ - an optimized network for lesion segmentation
\begin{enumerate}
    \item Initialize $\theta$ weights with pre-trained MobileNetV2, VGG16, or ResNet34 encoder
    \item Split dataset into 5 folds: $X_{1}, Y_{1}, ..., X_{5}, Y_{5}$

    \item \textbf{For each fold} $k$ $\leftarrow$ $1,2,3,4,5$

    \begin{enumerate}
    \item Define validation set as $X_v$, $Y_v$
    \item Define training set as $X_t$, $Y_t$
    \item Set up U-Net framework with pre-trained encoder, decoder, and Dice/Jaccard loss function
    \item Train network $F(x; \theta)$ on training set $(X_t, Y_t)$ for a specified number of epochs using Adam optimizer
    \item Evaluate network $F(x; \theta)$ on validation set $(X_v, Y_v)$ using Dice score and IoU as evaluation metrics
    \item Record evaluation metrics for validation set $(X_v, Y_v)$
    \item Preserve $\theta$ if evaluation metrics improve
    \end{enumerate}
    
    \item \textbf{return} $F(x; \theta)$ with best evaluation metrics
\end{enumerate}
\end{algorithm}

\subsection{Evaluation metrics for Classification}


The performance evaluation of benchmarking classification networks is typically conducted through quantitative assessment using widely recognized and significant metrics, including Pr, Re, F1 and Acc. The formulas for these metrics are outlined below:


Pr assesses the ratio of correctly identified positive samples and can be calculated using the following formula:
\begin{equation}
\\Pr=\frac{T_P}{F_P+T_P}\
\end{equation}
Re evaluates the ratio of correctly identified positive samples out of all the positive samples and can be computed using the following formula:

\begin{equation}
\\Re=\frac{T_P}{F_N+T_P}\
\end{equation}
The F1 integrates Pr and Re into a unified metric and is described as the harmonic mean of Re and Pr as follows:
\begin{equation}
\\F1\;score=\frac{T_P*2}{F_P+2*T_P+F_N}\
\end{equation}

Acc quantifies the ratio of accurately classified samples and can be computed using the following formula:
\begin{equation}
\\Acc.=\frac{T_N+T_P}{F_P+T_P+T_N+F_N}\
\end{equation}

where the number of true negatives, true positives, false negatives, and false positives are represented by  $T_N$, $T_P$, $F_N$, and $F_P$ respectively.
\subsection{Evaluation metrics for Segmentation}
The Dice score is a frequently utilized metric to evaluate the efficacy of a lesions segmentation of the fundus images algorithm by assessing the agreement between the predicted segmentation and its corresponding manually segmented ground truth on a pixel-wise basis. The Dice score can be determined using the following formula:

manual 
\begin{equation}
    \text{Dice score} = \frac{2 \times |\tilde{X} \cap \tilde{Y}|}{|\tilde{X}| + |\tilde{Y}|}
\end{equation}

The Intersection over Union (IoU) is a prevalent metric employed in numerous image segmentation algorithms to quantitatively assess their performance. It is calculated by dividing the intersection of two datasets, which are the ground truth and the algorithm's output, by their union. The IoU is typically computed using the following equation:
\begin{equation}
\text{IoU} = \frac{|\tilde{X} \cap \tilde{Y}|}{|\tilde{X} \cup \tilde{Y}|}
\end{equation}

Here, $\tilde{X}$ and $\tilde{Y}$ represent sets of binary values, and $|\tilde{X}|$ and $|\tilde{Y}|$ represent the total number of pixels in the ground truth and the segmentation network's output, respectively. The intersection $\tilde{X} \cap \tilde{Y}$ represents the number of pixels that are correctly segmented by both the ground truth and the model, and the union $\tilde{X} \cup \tilde{Y}$ represents the total number of pixels that are segmented by either the ground truth or the model (or both).

\section{Experimental results}  \label{mt6}
\subsection{Classification results}
In this study, we evaluated the performance of five SOTA pre-trained models, VGG16, DenseNet121, InceptionV3, MobileNetV2, and ResNet50, for a classification task. The models were evaluated using five-fold cross-validation (CV), and the metrics used to assess their performance were Acc, Pr, Re, and F1 Score.

The results showed that the MobileNetV2 DenseNet121 model achieved the highest mean Acc of 0.989 $\pm$0.006, closely followed by DenseNet121 with 0.985 $\pm$0.012. The VGG16 model achieved a mean Acc of 0.975 $\pm$0.030, while InceptionV3 achieved a mean Acc of 0.982 $\pm$0.008. The ResNet50 model had the lowest mean Acc of 0.972 $\pm$0.013.

In terms of Pr, DenseNet121 had the highest mean value of 0.986 $\pm$0.012, followed by MobileNetV2 with a mean value of 0.985±0.006. VGG16 and ResNet50 had similar mean Pr values of 0.976±0.029 and 0.976 $\pm$0.012, respectively. InceptionV3 had a mean Pr of 0.975 $\pm$0.009.

For Re, MobileNetV2 had the highest mean value of $0.993 $$\pm$$0.004$, followed by InceptionV3 with a mean value of $0.991 $$\pm$$0.015$. DenseNet121 had a mean Re of $0.986 $$\pm$$0.012$, while VGG16 and ResNet50 had similar mean Re values of $0.976 $$\pm$$0.03$ and $0.973  $$\pm$$0.013$, respectively.

For F1 Score, MobileNetV2 had the highest mean value of $0.989  $$\pm$$0.006$, followed by InceptionV3 with a mean value of $0.983  $$\pm$$0.005$. DenseNet121 had a mean F1 Score of $0.985  $$\pm$$0.012$, while VGG16 and ResNet50 had similar mean F1 Score values of $0.976 $$\pm$$0.03$ and $0.973 $$\pm$$0.013$, respectively.

Analyzing the standard deviation values, we observed that DenseNet121 and MobileNetV2 had the lowest variability in their performance, with a standard deviation (STD) of $1.186$ and $0.6$, respectively, for Acc. Similarly, for Pr, DenseNet121 had the lowest variability with an STD of $0.012$, followed by MobileNetV2 with an STD of $0.006$. For Re, MobileNetV2 had the lowest variability with an STD of $0.004$, followed by DenseNet121 with an STD of $0.012$. Finally, for F1 Score, MobileNetV2 had the lowest variability with an STD of $0.006$, followed by DenseNet121 with an STD of 0.012.

\begin{table}{}
\centering
\caption{Five-fold CV scores obtained for the benchmarking model.}
\label{tab:ten_fold}
\begin{tabular}{ccccc}
\hline
\multicolumn{5}{|c|}{\textbf{VGG16}} \\
\hline
\textbf{Fold} & \textbf{Acc} & \textbf{Pr} & \textbf{Re} & \textbf{F1 Score} \\ \hline
$1$ & $0.929$ & $0.931$ & $0.929$ & $0.929$ \\
$2$ & $0.951$ & $0.951$ & $0.951$ & $0.951$ \\
$3$ & $1.000$ & $1.000$ & $1.000$ & $1.000$ \\
$4$ & $1.000$ & $1.000$ & $1.000$ & $1.000$ \\
$5$ & $1.000$ & $1.000$ & $1.000$ & $1.000$ \\
\hline
\textbf{Avg.} & $0.976$ $\pm$ $0.030$ & $0.976$ $\pm$ $0.029$ & $0.976$ $\pm$ $0.030$ & $0.976$ $\pm$ $0.030$ \\
\hline
\multicolumn{5}{|c|}{\textbf{DenseNet121}} \\
\hline
\textbf{Fold} & \textbf{Acc} & \textbf{Pr} & \textbf{Re} & \textbf{F1 Score} \\
\hline
$1$ & $0.976$ & $0.978$ & $0.976$ & $0.976$ \\
$2$ & $0.976$ & $0.976$ & $0.976$ & $0.975$ \\
$3$ & $1.000$ & $1.000$ & $1.000$ & $1.000$ \\
$4$ & $1.000$ & $1.000$ & $1.000$ & $1.000$ \\
$5$ & $0.976$ & $0.977$ & $0.976$ & $0.976$ \\
\hline
\textbf{Avg.} & $0.985$ $\pm$ $0.012$ & $\textbf{0.986}$ $\pm$ $\textbf{0.011}$ & $0.986$ $\pm$ $0.012$ & $0.985$ $\pm$ $0.012$ \\
\hline
\multicolumn{5}{|c|}{\textbf{ResNet50}} \\
\hline
\textbf{Fold} & \textbf{Acc} & \textbf{Pr} & \textbf{Re} & \textbf{F1 Score} \\
\hline
$1$ & $0.967$ & $0.971$ & $0.967$ & $0.967$ \\
$2$ & $0.967$ & $0.968$ & $0.967$ & $0.966$ \\
$3$ & $0.966$ & $0.968$ & $0.966$ & $0.966$ \\
$4$ & $0.966$ & $0.971$ & $0.966$ & $0.967$ \\
$5$ & $1.000$ & $1.000$ & $1.000$ & $1.000$ \\
\hline
\textbf{Avg.} & $0.973$ $\pm$ $0.013$ & $0.976$ $\pm$ $0.012$ & $0.973$ $\pm$ $0.013$ & $0.973$ $\pm$ $\textbf{0.013}$ \\
\hline
\multicolumn{5}{|c|}{\textbf{InceptionV3}} \\
\hline
\textbf{Fold} & \textbf{Acc} & \textbf{Pr} & \textbf{Re} & \textbf{F1 Score} \\
\hline
$1$ & $0.980$ & $0.967$ & $0.993$ & $0.980$ \\
$2$ & $0.990$ & $0.978$ & $1.000$ & $0.989$ \\
$3$ & $0.970$ & $0.963$ & $0.993$ & $0.978$ \\
$4$ & $0.980$ & $0.988$ & $0.967$ & $0.977$ \\
$5$ & $0.990$ & $0.979$ & $1.000$ & $0.989$ \\
\hline
\textbf{Avg.} & $0.982$ $\pm$ $0.008$ & $0.975$ $\pm$ $0.009$ & $0.991$ $\pm$ $0.015$ & $0.983$ $\pm$ $0.005$ \\
\hline
\multicolumn{5}{|c|}{\textbf{MobileNetV2}} \\
\hline
\textbf{Fold} & \textbf{Acc} & \textbf{Pr} & \textbf{Re} & \textbf{F1 Score} \\
\hline
$1$ & $0.990$ & $0.987$ & $0.993$ & $0.990$ \\
$2$ & $0.995$ & $0.993$ & $0.997$ & $0.995$ \\
$3$ & $0.980$ & $0.975$ & $0.986$ & $0.980$ \\
$4$ & $0.985$ & $0.980$ & $0.990$ & $0.985$ \\
$5$ & $0.995$ & $0.992$ & $0.998$ & $0.995$ \\
\hline
\textbf{Avg.} & $\textbf{0.989}$ $\pm$ $\textbf{0.006}$ & $0.985$ $\pm$ $0.006$ & $\textbf{0.993}$ $\pm$ $\textbf{0.004}$ & $\textbf{0.989}$ $\pm$ $\textbf{0.006}$ \\

\hline
\end{tabular}
\end{table}

\subsection{Segmentation results}
In this article, we investigated various modified U-Net architectures for segmentation tasks, utilizing MobileNetV2, VGG16, InceptionV3, and ResNet34 as encoders within the U-Net framework. The weights of these models, pre-trained on the ImageNet dataset, are employed for $\mathscr{N}$ purposes. We employed a rigorous 5-fold CV methodology for model training and assessed performance based on Acc., Dice score, and IoU score.


    


\begin{figure*}[h]
  \centering
  \includegraphics[trim=0 0 0 0,clip,width=1\linewidth]{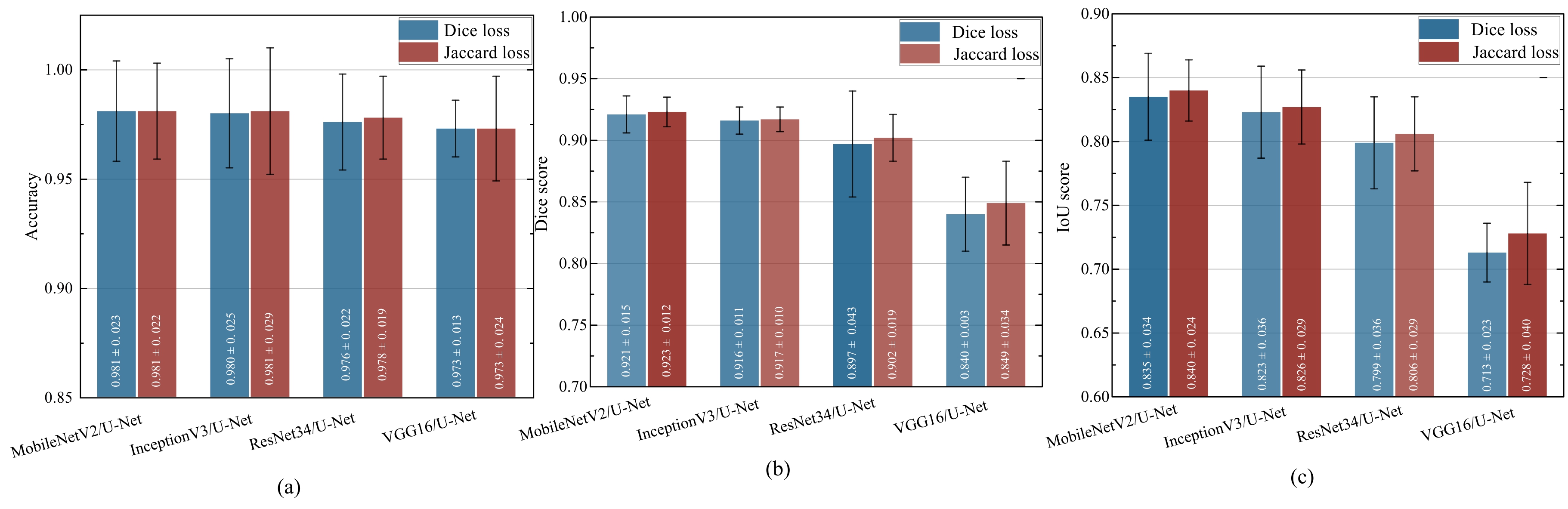}
  \vspace{-5mm}
  \caption{\textbf {(a) Acc, (b) Dice score, and (c) IoU score of different pre-trained modified U-Net models. }} 
  \label{seg_res}
\end{figure*}

In Fig.~\ref{seg_res}, the average performance of various models is demonstrated through a 5-fold CV process. Each model achieved an Acc higher than 0.97. The best-performing model, MobileNetV2/U-Net, recorded an average Acc of 0.981 with an STD of 0.023 and 0.022 for Dice and Jaccard losses, respectively. InceptionV3/U-Net, the second best model showed an accuraacy of 0.981$\pm$0.029 and 0.981$\pm$0.025 for the Jaccard and Dice losses, respectively. ResNet34/U-Net displayed mean Acc of 0.976$\pm$0.022 and 0.978$\pm$0.029 for the two losses. The VGG16 model exhibited the lowest mean Acc of 0.973$\pm$0.013 and 0.981$\pm$0.024.

In terms of the Dice score, MobileNetV2/U-Net outperforms the rest of the models. With Dice and Jaccard losses, it achieved mean Dice scores of 0.921$\pm$0.015 and 0.923$\pm$0.012, respectively. InceptionV3/U-Net, with the Dice loss showed a Dice score of 0.916$\pm$0.011 and and with Jaccard loss showed a Dice loss of 0.917$\pm$0.010. ResNet34/U-Net demonstrated a Dice score of  0.897$\pm$0.043 and 0.902$\pm$0.019 for the same losses. VGG16 demonstrated a mean Dice score of 0.840$\pm$0.003 and 0.849$\pm$0.034 for Dice and Jaccard losses respectively.
Additionally, the IoU score for MobileNetV2/U-Net was 0.835$\pm$0.034 and 0.840$\pm$0.024 for Dice and Jaccard losses, respectively. the second-best model, InceptionV3/U-Net demonstrated an IoU score of 0.823$\pm$0.036 and 0.826$\pm$0.029, respectively. ResNet34/U-Net displayed a mean IoU  score of 0.799$\pm$0.036 and 0.806$\pm$0.029 for the two losses. VGG16, on the other hand, had mean IoU scores of 0.713$\pm$0.023 and 0.728$\pm$0.040. Fig~\ref{seg_qualitative} illustrates the predicted masks generated by the different models.
It is worth noting that all the models were trained on binary cross-entropy loss  function as well which did not produce any satisfactory results.

\begin{table}[h]
\caption{Inference time and number of parameters of segmentation models}
\centering
\label{tab:seg_inf_parms}
\begin{tabular}{@{}ccc@{}}
\toprule
\textbf{Model}         & \textbf{Inference time (seconds)} & \textbf{Number of parameters} \\
\midrule \midrule
MobileNetV2/U-Net & $0.953$                    & $\textbf{8,047,441}$            \\
InceptionV3/U-Net    & $0.882$                   & $29,933,105$        \\
ResNet34/U-Net    & $0.849$                   & $24,456,154$         \\
VGG16/U-Net       & $\textbf{0.633}$                     & $23,752,273$           \\
\bottomrule
\end{tabular}
\end{table}


\begin{figure}[t]
    \centering
    \includegraphics[width=0.85\linewidth]{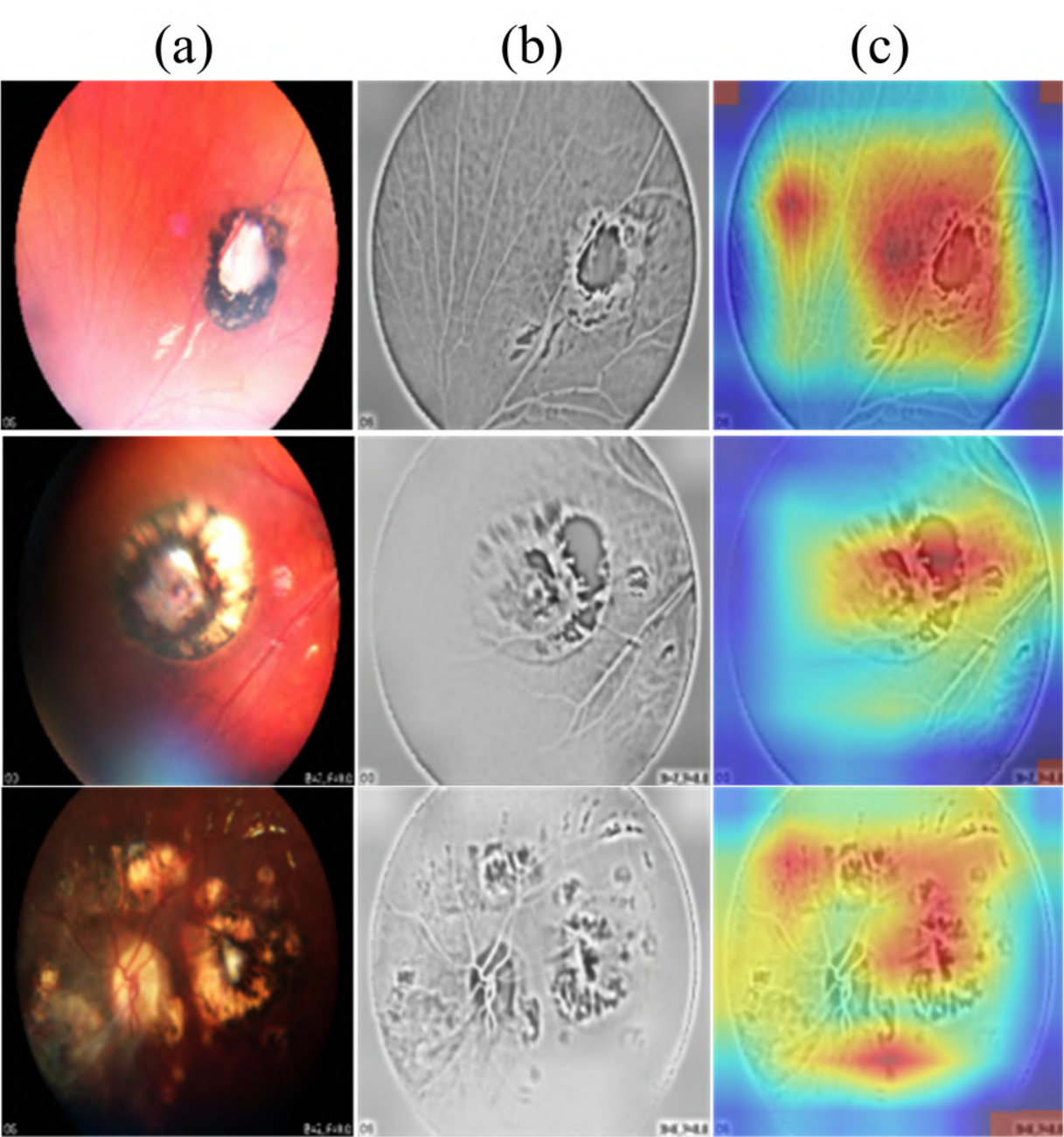}
    \caption{\textbf {(a) Original fundus image with infection region(first column), (b) preprocessed fundus image and (c) Grad-CAM visualizations for best performing MobileNetV2. 
}}
    \label{fig_gradcam}
\end{figure}

\section{Discussion} \label{mt7}
The quantitative classification results (shown in Tab.~\ref{tab:ten_fold}) indicate that we can achieve consistent identification of toxoplasmosis leveraging transfer learning. To gain a better qualitative insight, the final logit maps of a few samples correctly classified by the best performing MobileNetV2 are visualized in Fig.~\ref{fig_gradcam} leveraging the gradient-weighted class activation mapping (Grad-CAM) technique~\cite{selvaraju2017grad}. We can see that the model is indeed looking at and activating around infected regions in the fundus image visually validating our expectation. 

\begin{figure}[t]
    \centering
    \scalebox{.5}{
    \includegraphics[width=0.85\linewidth]{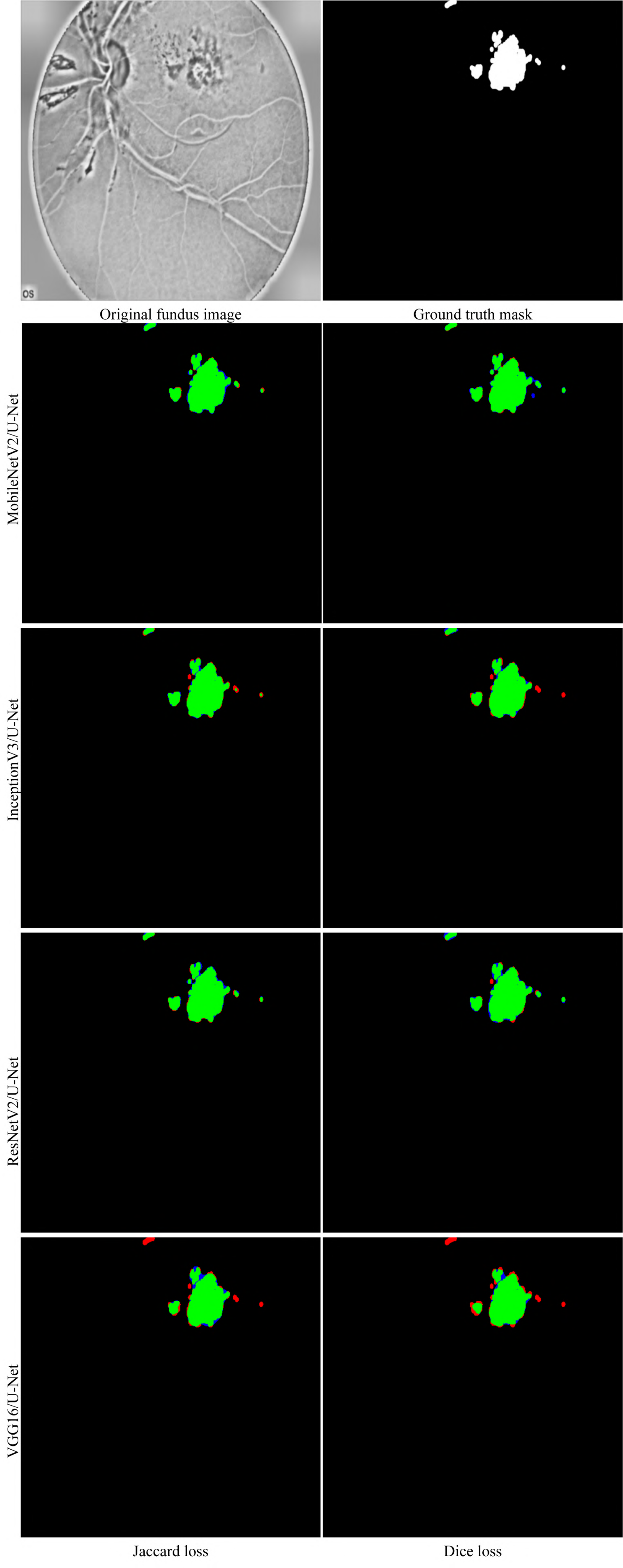}
        
    }
    \caption{\textbf{Visualization of the performance of segmentation networks in lesion segmentation. In this visualization, the true positive ($T_P$) region is represented by \textcolor{green}{'green'}, the false positive region by \textcolor{blue}{'blue'}, and the false negative regions by \textcolor{red}{'red'}.}}
    \label{seg_qualitative}
\end{figure}

\begin{figure*}[t]
    \centering
    \scalebox{.9}{
        \includegraphics[width=0.85\linewidth]{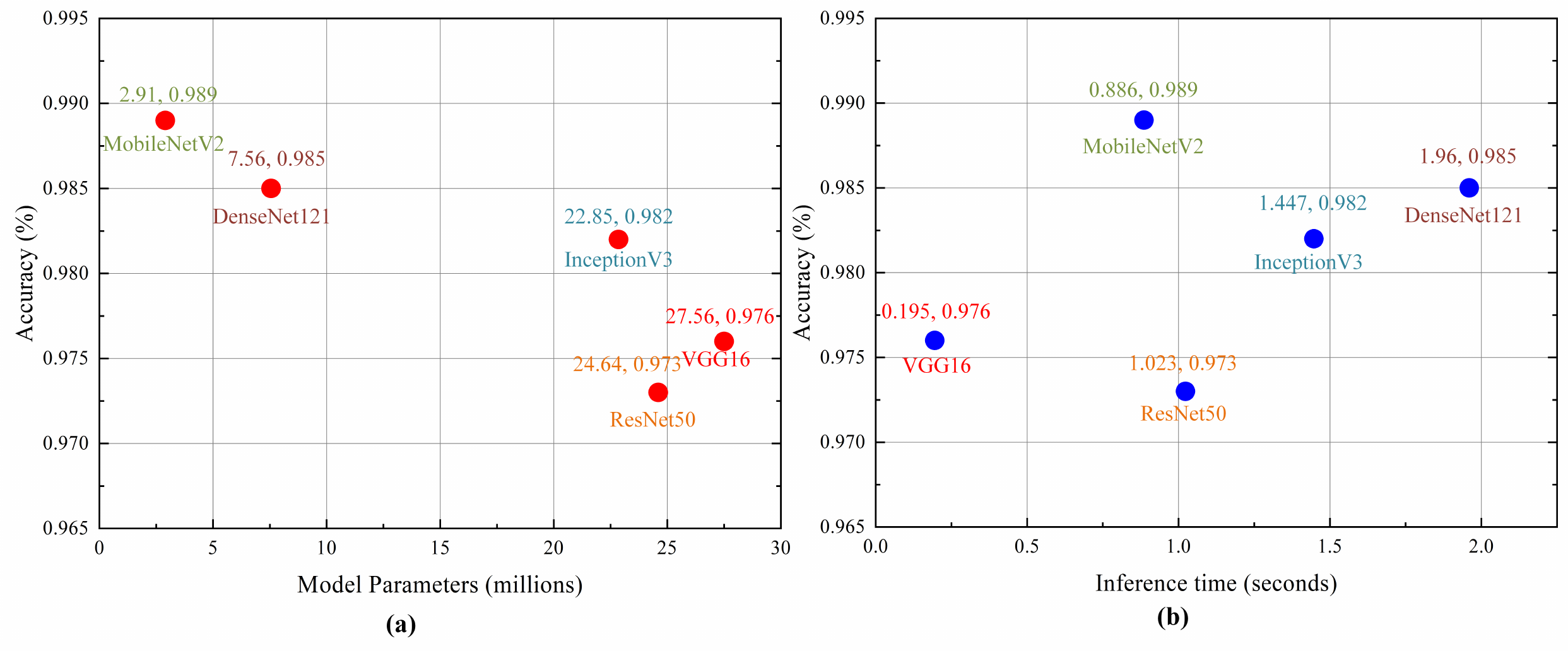}
    }
    
    \caption{\textbf {(a) Acc vs Model Parameters (M), (b) Acc vs Inference Time (sec)}}
    \label{time}
\end{figure*} 
\par 
From Fig.~\ref{seg_res}, it is evident that the MobileNetV2/U-Net model outperforms the rest three models i.e., InceptionV3/U-Net, ResNet34/U-Net, and VGG16/U-Net across all the performance metrics. However, when Jaccard loss is employed as the loss function, it shows a marginal improvement over the results obtained with Dice loss. This phenomenon is also reflected in Fig.~\ref{seg_qualitative}, which assesses the segmentation results in a qualitative manner. 
It can be clearly seen that the MobileNetV2/U-Net model yielded remarkable segmentation performance, that is nearly identical to the ground
truth with much less mis-segmented tissue. In contrast, VGG16/U-Net gives unsatisfactory results, where the fine details at the edges could be segmented correctly. Although InceptionV3/U-Net and ResNet34/U-Net improves the results, but the performance is still not promising. The success of
MobileNetV2/U-Net can be attributed to its depthwise separable convolution-based encoder architecture, which splits the standard convolution into two separate operations: a depthwise convolution (applies a single filter to each input channel) and a pointwise convolution (applies a 1x1 filter to combine the output of the depthwise convolution)~\cite{sandler2018mobilenetv2}. Additionally, MobileNet uses global average pooling instead of fully connected layers. These optimizations lead to enhanced speed and memory efficiency when compared to conventional CNNs.
\par
As demonstrated in Fig.~\ref{time}, we have analyzed the architectures, VGG16, ResNet50, DenseNet121, InceptionV3 and MobileNetV2 for the classification task in terms of Acc, inference time and  the number of parameters to assess their complexity. Our experimental results show that MobileNetV2 achieved the highest Acc among the 5 models, despite having the least number of parameters. DenseNet121 and InceptionV3 also performed competitively but they had approximately 2.6x and 7.85x less parameters, respectively compared to the best performing MobileNetV2. VGG16 achieved the second lowest Acc although it has around 9.5x parameters than the best performing MobileNetV2. However, it had the lowest inference time (0.195 sec). MobileNetV2 also appeared to be significantly faster than the other models with a inference time of 0.886 sec. Although DenseNet121 competitively performed in terms of Acc, having the highest inference time of 1.96 sec, it is not suitable for implementing in edge devices where faster inference is crucial.  Therefore, our results suggest that MobileNetV2 is a promising architecture for classifying OT from retinal fundus images due to its high Acc, low inference time, and low computational complexity. Similar trend is observed for the segmentation task in Table.~\ref{tab:seg_inf_parms}. MobileNetV2/U-Net performs the best despite having the least parameters, but it also has a drawback of compartaively greater inference time (0.953 sec). Again, VGG16/U-Net works with the greatest speed with a inference time of 0.633 sec but does not perform in a satisfactory manner despite having almost 3x parameters than the best performing MobileNetV2/U-Net.

\section{Limitation and Future Work}\label{mt9}
Although our experiments have yielded promising results, it is important to acknowledge that the dataset used in our study is limited in size. This constraint not only limits the system's reliability when faced with atypical cases and images but also restricts the development of complex and deep neural network architectures. As a result, we had to rely on pre-trained networks instead of intricate, task-specific architectures.

To address this limitation, our future work aims to collect more fundus images to create a larger dataset. We plan to collect these images in a manner that maintains a relatively even ratio between healthy and unhealthy (\textit{T. gondii} positive) patient images, as well as an adequate amount of active, inactive, and active/inactive images for future classification and segmentation experiments. In addition, we plan to leverage existing large fundus image datasets from other domains and use them as part of a self-supervised learning algorithm to improve the resulting solution's performance. Furthermore, we plan on deploying more advanced architecture models, such as the image transformer classifier, to examine the effectiveness of SOTA architectures on our dataset.
 
\section{Conclusion}  \label{mt10}
This paper presents a comprehensive benchmarking of transfer learning approaches for medical image segmentation, classification, and transfer learning, with a particular focus on the OT diagnosis community. Through an evaluation of various preprocessing techniques for OT region diagnosis and segmentation, we have successfully employed pre-trained networks including VGG16, InceptionV3, ResNet, DenseNet121, and MobileNetV2. Remarkably, the model with the backbone of pre-trained MobileNetV2 has shown exceptional performance in both classification and segmentation tasks, outperforming all other models with substantial margins. The promising experimental results provide strong evidence for the reliability and robustness of diagnosis and semantic segmentation for OT lesions. This study has the potential to open up new avenues for advanced medical image analysis and automated diagnosis in the field of OT.
\bibliographystyle{IEEEtran}
\bibliography{ref}

\end{document}